\documentclass[aps, prb, 10pt, twocolumn, superscriptaddress]{revtex4-2}

\usepackage[english]{babel}
\usepackage[utf8]{inputenc}
\usepackage{fourier}
\usepackage[T1]{fontenc}

\usepackage{acronym}
\usepackage{bm}
\usepackage{mathtools}
\usepackage{microtype}
\usepackage{tikz}

\usetikzlibrary{decorations.markings}
\usetikzlibrary{decorations.pathmorphing}

\usepackage[colorlinks, allcolors=purple]{hyperref}
\usepackage[figure, figure*]{hypcap}

\let\Re\relax
\let\Im\relax
\DeclareMathOperator\Re{Re}
\DeclareMathOperator\Im{Im}

\def\D{\mathrm d}
\def\E{\mathrm e}
\def\I{\mathrm i}

\def\abs#1{|#1|}
\def\bra#1{\langle#1|}
\def\ket#1{|#1\rangle}
\def\bracket#1#2{\langle#1|#2\rangle}
\def\av#1{\langle#1\rangle}
\def\sub#1{_{\textnormal{#1}}}
\def\super#1{^{\textnormal{#1}}}

\def\ph{^{\smash{\textnormal{ph}}}}

\let\Pi\varPi
\let\Sigma\varSigma
\let\epsilon\varepsilon
\let\mat\bm
\let\s\textsubscript
\let\tilde\widetilde
\let\vec\bm

\tikzset{
    el/.style={
        decoration={markings, mark=at position 1/2 with {\arrow[xshift=1pt, scale=1.5]>}},
        preaction={decorate},
    },
    ph/.style={
        decoration={coil, pre length=1.5mm, segment length=2mm, post length=1.5mm, amplitude=2mm},
        decorate,
    },
    elel/.style={
        decoration={snake, pre length=1mm, segment length=2mm, post length=1mm},
        decorate,
    },
}

\newacro{ARPES}{angle-resolved photoemission spectroscopy}
\newacro{DC}{direct current}
\newacro{DFG}{Deutsche Forschungsgemeinschaft}
\newacro{DFPT}{density-functional perturbation theory}
\newacro{DFT}{density-functional theory}
\newacro{DMFT}{dynamical mean-field theory}
\newacro{DOS}{density of states}
\newacro{FNRS}[F.R.S.-FNRS]{Fonds de la Recherche Scientifique - FNRS}
\newacro{HWHM}{half width at half maximum}
\newacro{PBE}{Perdew--Burke--Ernzerhof}
\newacro{QPA}{quasiparticle approximation}
\newacro{RPA}{random-phase approximation}
\newacro{RTA}{relaxation-time approximation}
\newacro{SOC}{spin--orbit coupling}
\newacro{SOP}{sum over poles}
\newacroplural{SOP}[SOPs]{sums over poles}

\newlength\eX
\def\orcid#1{\,\href{https://orcid.org/#1}{\includegraphics[height=\eX]{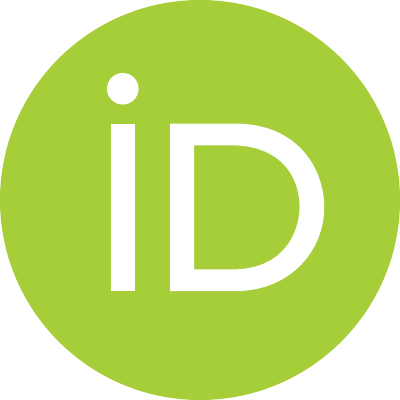}}}

\def\MAPEX{%
    U~Bremen Excellence Chair,
    Bremen Center for Computational Materials Science,
    and MAPEX Center for Materials and Processes,
    University of Bremen,
    28359 Bremen,
    Germany}

\def\HMI{%
    Hybrid Materials Interfaces Group,
    Faculty of Production Engineering,
    University of Bremen,
    28359 Bremen,
    Germany}

\def\Louvain{%
    European Theoretical Spectroscopy Facility,
    Institute of Condensed Matter and Nanosciences,
    Universit\'e catholique de Louvain,
    1348 Louvain-la-Neuve,
    Belgium}

\def\WEL{%
    WEL Research Institute,
    1300 Wavre,
    Belgium}

\def\EPFL{%
    Theory and Simulation of Materials~(THEOS),
    and National Centre for Computational Design and Discovery of Novel Materials~(MARVEL),
    \'Ecole Polytechnique F\'ed\'erale de Lausanne,
    1015 Lausanne,
    Switzerland}

\def\Cambridge{%
    Theory of Condensed Matter,
    Cavendish Laboratory,
    University of Cambridge,
    Cambridge CB3 0US,
    United Kingdom}

\def\Caltech{%
    Department of Applied Physics and Materials Science,
    California Institute of Technology,
    Pasadena CA 91125,
    USA}

\begin{document}

\title{Unified treatment of local dynamical interactions in correlated metals using Eliashberg theory}

\author{Jan Berges\orcid{0000-0001-7041-2906}}
\affiliation\MAPEX
\affiliation\HMI

\author{Samuel Ponc\'e\orcid{0000-0003-1159-8389}}
\affiliation\Louvain
\affiliation\WEL

\author{Mario Caserta\orcid{0000-0003-2812-9386}}
\affiliation\EPFL

\author{Nicola Marzari\orcid{0000-0002-9764-0199}}
\affiliation\MAPEX
\affiliation\EPFL
\affiliation\Cambridge

\author{Tommaso Chiarotti\orcid{0000-0003-1572-4987}}
\affiliation\Caltech

\begin{abstract}
    First-principles studies of materials with strong electronic correlations introduce additional theoretical complexities and computational costs, even more so if lattice degrees of freedom also play a role.
    To address these challenges, we show how direct and phonon-mediated electron--electron interactions can be treated consistently.
    The key observation is that the $GW$ and Fan--Migdal self-energies are topologically the same diagram, so that they can be unified using a single effective interaction, which is the sum of the two interactions.
    The Dyson equation for the electron Green function then yields normal-state Eliashberg equations, which can be viewed as a subset of the Hedin--Baym equations for coupled electrons and phonons [\href{https://doi.org/10.1103/PhysRevX.13.031026}{Phys.~Rev.~X~\textbf{13}, 031026 (2023)}].
    We solve these dynamical equations in a local approximation to calculate from first principles the temperature-dependent quasiparticle spectrum, with close attention to the low-energy window determining transport.
    We apply this extended dynamical Hubbard approach to the correlated metal Sr\s2RuO\s4, which has long served as a benchmark system for advanced electronic-structure methods.
    The resulting renormalization of quasiparticle bands is in qualitative agreement with experiments and results from \acl*{DMFT}.
    However, here the low-energy linewidth and associated resistivity within the Green--Kubo formalism in the bubble approximation are underestimated, suggesting the need for vertex corrections.
    This work highlights how simple dynamical formulations can provide an extensible framework for studying correlated materials, which treats electronic and vibrational excitations on equal footing.
\end{abstract}

\maketitle

\section{Introduction}

In the past decades, \ac{DFT} \cite{Hohenberg1964} has become the state-of-the-art tool to efficiently simulate materials at the nanoscale.
In correlated systems, where interactions qualitatively reshape low-energy excitations and the effective single-particle picture breaks down, more advanced methods such as \ac{DMFT} \cite{Georges1996, Kotliar2006, Abramovitch2025} are required, where data from \ac{DFT} is often a good starting point.
In some cases, one may resort to computationally less intensive methods such as multi-orbital iterative perturbation theory \cite{Gurshidali2026} or Hubbard-corrected ``\ac{DFT}$+U$'', which alleviates self-interaction errors in \ac{DFT} \cite{Liechtenstein1995, Dudarev1998}.
Recently, \emph{dynamical} Hubbard functionals ``\ac{DFT}$+W(\omega)$'' have been successfully used to obtain spectral and thermodynamic properties of correlated materials \cite{Vanzini2023, Chiarotti2022, Chiarotti2023, Chiarotti2024, Chiarotti2025, Caserta2026}.
Here, the electronic structure from \ac{DFT} is complemented with a $GW$-like self-energy with a Green function $G$ and a screened interaction $W$ from the \ac{RPA}, both localized in the correlated orbitals.

In this work, we use such dynamical corrections to obtain the spectral and transport properties of the unconventional superconductor and correlated metal Sr\s2RuO\s4 \cite{Maeno1994, Mackenzie2003, Bergemann2003, Takahashi2014, Kim2018, Zingl2019, Li2022, Noad2023, Maeno2024, Krsnik2026}, which provides an exceptionally clean and experimentally accessible platform to study the interplay between Fermi-liquid quasiparticles and strong correlations.
In addition to effects from direct electron--electron interaction, we include phonon-mediated interaction on equal footing, replacing $W$ by an effective $\tilde W$.
More precisely, we complement the local $GW$ diagram with a local Fan--Migdal self-energy based on electron--phonon matrix elements from \ac{DFPT} \cite{Gonze1997, Baroni2001}.
A similar study has already been carried out using \ac{DMFT} \cite{Abramovitch2023}, providing valuable reference data.

Existing dynamical Hubbard studies rely on a \ac{SOP} representation of \emph{time-ordered} propagators \cite{Engel1993, Leon2021, Ferretti2024}, which turns convolutions into efficient analytical operations and allows for solving the Dyson equation via algorithmic inversion \cite{Chiarotti2022, Chiarotti2023, Chiarotti2024, Chiarotti2025}.
While this approach is ideally suited to describe the overall renormalization of the bands, the time ordering and limited basis size complicate an accurate description of the temperature-dependent low-energy Green and spectral functions (see Appendix~\ref{app:matsubara}), which are crucial for determining the electrical resistivity.
Therefore, we instead use imaginary- and real-axis formulations, where the \emph{retarded} self-energy is obtained from analytic continuation \cite{Vidberg1977, Lee1996} or directly on a logarithmic energy grid.
Compared to previous works \cite{Chiarotti2024, Chiarotti2025, Caserta2026}, we limit the description to the low-energy bands and neglect changes to the Kohn--Sham potential due to charge renormalization.
We show that such a local dynamical $GW$ \cite{vonBarth1996, Holm1998, Benedict2002} is formally equivalent \cite{Spataru2026} to the orbital-resolved local Eliashberg theory in the normal state.
We therefore use the software \textsc{ebmb} \cite{Berges2016b} developed to solve the band- or orbital-resolved local Eliashberg equations.
Here, the double-counting term used in the dynamical Hubbard approach \cite{Chiarotti2024} is already implicitly considered because of different conventions for the static exchange.

Using one-shot local $G_0 \tilde W_0$, we qualitatively reproduce the spectral function from \ac{ARPES} \cite{Tamai2019}, where the electronic bandwidth is reduced compared to \ac{DFT}, showing that \ac{DFT}$+W(\omega)$ can bridge the gap between \ac{DFT} and \ac{DMFT} in this material.
Knowledge of the low-energy spectral function allows us to compute the electrical resistivity using the Green--Kubo approach in the bubble approximation \cite{Lihm2026}.
However, while the phonon-mediated interaction leads to a resistivity contribution in agreement with previous results \cite{Abramovitch2023}, the influence of the direct electron--electron interaction is underestimated compared to \ac{DMFT}.
This is because the dynamical part of the interaction in \ac{DFT}$+W(\omega)$, which involves nonlocal screening \cite{Caserta2026} and is usually ignored in \ac{DMFT}, acts at high energies and consequently only slightly broadens the low-energy states.

This paper is organized as follows:
In Sec.~\ref{sec:methods}, we present the Green-function formalism underlying dynamical Hubbard and Eliashberg theory and the calculation of spectral and transport properties.
In Sec.~\ref{sec:implementation}, we list computational details.
In Sec.~\ref{sec:results}, we show our results for Sr\s2RuO\s4.
Finally, in Sec.~\ref{sec:conclusions}, we summarize our findings and discuss further perspectives.

\section{Dynamical Hubbard and phonon-mediated interactions on equal footing}
\label{sec:methods}

\subsection{Local \textit{GW} and Fan--Migdal approximation}

We consider local (momentum independent) and dynamical (energy dependent) propagators within the correlated subspace spanned by the orbitals $\alpha, \beta$.
We denote fermionic and bosonic energies with $\epsilon$ and $\omega$.
Since the diagrammatic expressions of the dynamical Hubbard self-energy (that is, local $GW_0$) and of local Fan--Migdal theory coincide, combining them only amounts to considering an effective electron--electron interaction $\tilde W_{\alpha \beta}(\omega)$ given by the sum of the screened Coulomb interaction $W_{\alpha \beta}(\omega)$ and the phonon-mediated interaction $W \ph_{\alpha \beta}(\omega)$,
\begin{equation}
\begin{tikzpicture}[baseline=-1mm, thick, double distance=1pt, line cap=butt]
    \draw [densely dashed, double] (0, 0) -- (2, 0);
    \node [below=1mm] at (1, 0) {$\tilde W_{\alpha \beta}(\omega)$};
    \node at (2.25, 0) {$=$};
    \draw [elel, double] (2.5, 0) -- (4.5, 0);
    \node [below=1.5mm] at (3.5, 0) {$W_{\alpha \beta}(\omega)$};
    \node at (4.75, 0) {$+$};
    \draw [ph, double] (5, 0) -- (7, 0);
    \node [below=2.5mm] at (6, 0) {$W \ph_{\alpha \beta}(\omega)$};
    \draw [fill=white] (5, 0) circle (0.8mm);
    \draw [fill=white] (7, 0) circle (0.8mm);
\end{tikzpicture}.
\end{equation}
The bare Green function $G_{0 \alpha}(\epsilon)$ and the screened Green function $G_\alpha(\epsilon) \equiv G_{\infty \alpha}(\epsilon)$ fulfill the Dyson equation
\begin{equation}
\begin{tikzpicture}[baseline=-1mm, thick, double distance=1pt, line cap=butt]
    \draw [el, double] (0, 0) -- (1, 0);
    \node [below=1mm] at (0.5, 0) {$G_\alpha(\epsilon)$};
    \node at (1.25, 0) {$=$};
    \draw [el] (1.5, 0) -- (2.5, 0);
    \node [below=1mm] at (2, 0) {$G_{0 \alpha}(\epsilon)$};
    \node at (2.75, 0) {$+$};
    \draw [el] (3, 0) -- (4, 0);
    \node [below=1mm] at (3.5, 0) {$G_{0 \alpha}(\epsilon)$};
    \draw [el, double] (4, 0) -- (6, 0);
    \node [below=1mm] at (5, 0) {$G_\beta(\epsilon + \omega)$};
    \draw [densely dashed, double] (4, 0) arc (180:0:1);
    \node [below=2mm] at (5, 1) {$\tilde W_{\alpha \beta}(\omega)$};
    \draw [el, double] (6, 0) -- (7, 0);
    \node [below=1mm] at (6.5, 0) {$G_\alpha(\epsilon)$};
    \fill (4, 0) circle (0.6mm);
    \fill (6, 0) circle (0.6mm);
\end{tikzpicture}.
\end{equation}
Since they are diagonal in the orbital basis, we only consider the intra-orbital density--density terms ($\alpha = \beta$) and the inter-orbital exchange terms ($\alpha \neq \beta$) of our Coulomb interaction from \ac{RPA}, while inter-orbital density--density interactions do not contribute.
The phonon-mediated interaction inherits the orbital dependence of the electron--phonon matrix elements from \ac{DFPT}.

The above diagrams constitute an approximation to a subset of the Hedin--Baym equations governing the interplay of interacting electrons and phonons \cite{Stefanucci2023}.
Among others, the static Debye--Waller diagram \cite{Ponce2025} is omitted here.

The bare Green function is defined by the orbital-resolved electronic \ac{DOS} $N_\alpha(\epsilon)$ from \ac{DFT},
\begin{subequations}
\label{eq:G0}
\begin{align}
    G_{0 \alpha}(\epsilon) &= \int_{-\infty}^\infty \D \epsilon' \frac{N_\alpha(\epsilon')}{\epsilon - \epsilon' + \mu_0},
    \\
    \label{eq:DOS}
    N_\alpha(\epsilon) &= \frac 1 {N_{\vec k}} \sum_{\vec k n} \abs{\bracket{\vec k \alpha}{\vec k n}}^2 \delta(\epsilon_{\vec k n} - \epsilon),
\end{align}
\end{subequations}
where $\epsilon_{\vec k n}$ is the Kohn--Sham eigenenergy for momentum $\vec k$ and spin-degenerate band $n$, $\mu_0$ is the noninteracting chemical potential, $N_{\vec k}$ is the number of $\vec k$ points, $\ket{\vec k n}$ is an eigenstate, and $\ket{\vec k \alpha}$ a Wannier basis state.
Similarly, the screened phonon-mediated interaction is defined by its spectral function $B \ph_{\alpha \beta}(\omega)$, which consists of the phonon \ac{DOS} weighted with the absolute square of the electron--phonon coupling strength from \ac{DFPT},
\begin{subequations}
\label{eq:Wph}
\begin{align}
    W \ph_{\alpha \beta}(\omega) &= \int_0^\infty \D \omega' \frac{2 \omega' B \ph_{\alpha \beta}(\omega')}{\omega^2 - \omega'^2},
    \\
    \label{eq:Bph}
    B \ph_{\alpha \beta}(\omega) &= \frac 1 {N_{\vec q} N_{\vec k}} \sum_{\pm \vec q \nu \vec k} (\pm) \abs{g_{\vec q \nu \vec k \alpha \beta}}^2 \delta(\pm \omega_{\vec q \nu} - \omega),
\end{align}
\end{subequations}
where $\omega_{\vec q \nu}$ is the phonon energy for momentum $\vec q$ and mode $\nu$.
Here, the phonon propagator $D_{\vec q \nu}(\omega) = 2 \omega_{\vec q \nu} / (\omega^2 - \omega_{\vec q \nu}^2)$ is bracketed with the electron--phonon matrix elements \footnote{Note that $\hbar$ does not appear under the square root since $\omega_{\vec q \nu}$ is an energy instead of a frequency here.}
\begin{align}
    g_{\vec q \nu \vec k \alpha \beta} = \frac \hbar {\sqrt{2 \omega_{\vec q \nu}}} \sum_{\kappa x} e_{\vec q \kappa x \nu} \frac 1 {\sqrt{M_\kappa}} \bra{\vec k {+} \vec q \alpha} \frac{\partial V \super{SCF}}{\partial u_{\vec q \kappa x}} \ket{\vec k \beta},
\end{align}
where $e_{\vec q \kappa x \nu}$ is an eigenvector of the dynamical matrix, $M_\kappa$ is the mass of the $\kappa$th atom, $V \super{SCF}$ is the self-consistent potential, and $u_{\vec q \kappa x}$ is a periodic displacement in the $x$th Cartesian direction.
The screened Coulomb interaction from \ac{RPA} is defined not only by its spectral function $B_{\alpha \beta}(\omega)$ but also by a real offset given by the bare interaction $v_{\alpha \beta}$ \cite{vonBarth1996, Holm1998, Benedict2002},
\begin{subequations}
\label{eq:Wel}
\begin{align}
    W_{\alpha \beta}(\omega) &= \int_0^\infty \D \omega' \frac{2 \omega' B_{\alpha \beta}(\omega')}{\omega^2 - \omega'^2} + v_{\alpha \beta},
    \\
    \label{eq:Bel}
    B_{\alpha \beta}(\omega) &= -\frac 1 \pi \Im W_{\alpha \beta}(\omega + \I \eta),
\end{align}
\end{subequations}
where $\eta$ is a positive infinitesimal (in practice, small but finite) broadening.
Adding up Eqs.~\eqref{eq:Wph} and \eqref{eq:Wel} yields the effective electron--electron interaction
\begin{subequations}
\label{eq:W}
\begin{align}
    \tilde W_{\alpha \beta}(\omega) &= \int_0^\infty \D \omega' \frac{2 \omega' \tilde B_{\alpha \beta}(\omega')}{\omega^2 - \omega'^2} + v_{\alpha \beta},
    \\
    \label{eq:B}
    \tilde B_{\alpha \beta}(\omega) &= B_{\alpha \beta}(\omega) + B \ph_{\alpha \beta}(\omega).
\end{align}
\end{subequations}
After the $i$th iteration towards self-consistency, the screened Green function $G_{i \alpha}(\epsilon)$ and its spectral function $A_{i \alpha}(\epsilon)$ follow from the inversion of the Dyson equation,
\begin{subequations}
\label{eq:G}
\begin{align}
    G_{i \alpha}(\epsilon) &= \int_{-\infty}^\infty \D \epsilon' \frac{N_\alpha(\epsilon')}{\epsilon - \epsilon' + \mu_i - \Sigma_{i \alpha}(\epsilon)} \equiv \int_{-\infty}^\infty \D \epsilon' \frac{A_{i \alpha}(\epsilon')}{\epsilon - \epsilon'},
    \\
    \label{eq:A}
    A_{i \alpha}(\epsilon) &= -\frac 1 \pi \Im G_{i \alpha}(\epsilon + \I \eta),
\end{align}
\end{subequations}
where $\mu_i$ is the chemical potential.
Here, the local self-energy $\Sigma_{i \alpha}(\epsilon)$ can conveniently be obtained from the spectral functions $A_{i \alpha}(\epsilon)$, with $A_{0 \alpha}(\epsilon) = N_\alpha(\epsilon + \mu_0)$, and $\tilde B_{\alpha \beta}(\omega)$.
It reads
\begin{multline}
    \label{eq:Sigma}
    \Sigma_{i \alpha}(\epsilon) = \sum_\beta \int_{-\infty}^\infty \D \epsilon' \, A_{i - 1 \beta}(\epsilon') \Biggl[ v_{\alpha \beta} \bigl( \tfrac12 - f(\epsilon') \bigr)
    \\
    + \int_0^\infty \D \omega \, \tilde B_{\alpha \beta}(\omega) \sum_\pm (\pm 1) \frac{f(\epsilon') + n(\pm \omega)}{\epsilon - \epsilon' \pm \omega} \Biggr].
\end{multline}
The dependence on temperature $T$ enters only through the Fermi function $f(\epsilon) = 1 / (\E^{\epsilon / k \sub B T} + 1)$ and the Bose function $n(\omega) = 1 / (\E^{\omega / k \sub B T} - 1)$, as derived in Appendix~\ref{app:matsubara}.
$\Sigma_{1 \alpha}(\epsilon)$ corresponds to one-shot local $G_0 \tilde W_0$ and $\Sigma_{\infty \alpha}(\epsilon)$ to self-consistent local $G \tilde W_0$.

Note that the term with $\frac12 - f$ in the first line of Eq.~\eqref{eq:Sigma} is different from the $-f$ found in the $GW$ literature \cite{Benedict2002}.
It is minus the average of the imaginary-time free-fermion Green functions $G_0(\tau = 0^-) = f$ and $G_0(\tau = 0^+) = f - 1$ evaluated just below and above zero.
It stems from the Eliashberg theory, where the Matsubara summations are usually evaluated numerically on a finite number of energies centered around zero (see Appendix~\ref{app:matsubara}).
Conversely, in $GW$, the convolution of the Green function and the interaction is usually performed on the real axis, introducing a factor $\E^{\I \eta \omega}$ that allows for closing the integration contour despite the static part of the interaction \cite{Hedin1965}, giving rise to the usual static-exchange term.
Using $\frac12 - f$ renders a double-counting correction unnecessary, as the associated energy term $\propto (1 - N) N$ vanishes in the fully localized or atomic limit of orbital occupations $N \in \{0, 1\}$ \cite{Czyzyk1994, Jiang2010}.
Indeed, for $v_{\alpha \beta} = W_\infty \delta_{\alpha \beta}$ and since $\int \D \epsilon A_{i \alpha}(\epsilon) = 1$, the term $(\frac12 - f) v_{\alpha \beta}$ corresponds to the static term in the dynamical Hubbard functional and the \ac{DFT}$+U$ formulation by \citeauthor{Dudarev1998} \cite{Dudarev1998}.

Since the physical value of the imaginary part of the retarded self-energy can become very small, especially in the vicinity of the Fermi level, one has to make sure that $\eta \rightarrow 0$.
Using $\Im[1 / (\epsilon + \I 0^+)] = -\pi \delta(\epsilon)$, one obtains
\begin{multline}
    \label{eq:ImSigma}
    \Im \Sigma_{i \alpha}(\epsilon + \I 0^+) = -\pi \sum_\pm (\pm 1) \sum_\beta \int_0^\infty \D \omega \, A_{i - 1 \beta}(\epsilon \pm \omega) \tilde B_{\alpha \beta}(\omega)
    \\
    \times \bigl( f(\epsilon \pm \omega) + n(\pm \omega) \bigr).
\end{multline}
The computational cost of Eq.~\eqref{eq:ImSigma} is strongly reduced compared to Eq.~\eqref{eq:Sigma}, since there is only one energy integral instead of two.
The dynamical part of the full complex self-energy in Eq.~\eqref{eq:Sigma} is recovered using the Kramers--Kronig relation or spectral representation
\begin{equation}
    \label{eq:krakro}
    \Sigma_{i \alpha}(\epsilon) = -\frac 1 \pi \int_{-\infty}^\infty \D \epsilon' \frac{\Im \Sigma_{i \alpha}(\epsilon' + \I 0^+)}{\epsilon - \epsilon'} + \Sigma_{i \alpha}(\infty).
\end{equation}
The chemical potential $\mu_i$ hidden in $A_{i \beta}(\epsilon)$ [see Eq.~\eqref{eq:G}] is adjusted such that the electron number $n \sub e$ is conserved.
With a factor of two accounting for the spin degree of freedom,
\begin{equation}
    \label{eq:ne}
    n \sub e = 2 \int_{-\infty}^\infty \D \epsilon \, f(\epsilon) \sum_\beta A_{i \beta}(\epsilon).
\end{equation}
Note that we ignore the temperature dependence of $G_{0 \alpha}(\epsilon)$, $W \ph_{\alpha \beta}(\omega)$, and $W_{\alpha \beta}(\omega)$, which are computed only once for a fixed smearing.
While this is similar to an ``electronic temperature'', there is no one-to-one correspondence, as we are using a Marzari--Vanderbilt smearing \cite{Marzari1999} rather than Fermi--Dirac occupations.

\subsection{Local Eliashberg theory}

Another approach that is based on $GW$-like diagrams, neglecting vertex corrections of the electron--phonon interaction as per Migdal's theorem \cite{Migdal1958}, is the Eliashberg theory of superconductivity \cite{Eliashberg1960, Spataru2026}.
In fact, the only differences are that (i) an anomalous propagator describing the formation of Cooper pairs is considered in addition to the normal electron Green function and (ii) the Coulomb interaction is usually assumed to be static and only acts on the anomalous part, as it is effectively included in the noninteracting electronic bands and must not be double counted.

Here, we work on the imaginary axis and consider local, orbital-dependent functions of fermionic Matsubara energies $\epsilon_n = (2 n + 1) \pi k \sub B T$ with integer $n$.
In the Nambu formalism \cite{Nambu1960}, the self-energy becomes a $2 \times 2$ matrix with the normal and anomalous parts on the diagonal and off-diagonal.
In terms of the renormalization function $Z_{i \alpha}(\I \epsilon_n)$, the superconducting order parameter $\phi_{i \alpha}(\I \epsilon_n)$, and the energy shift $\chi_{i \alpha}(\I \epsilon_n)$, it reads
\begin{equation}
    \label{eq:Sigma2x2}
    \mat \Sigma_{i \alpha}(\I \epsilon_n) = \I \epsilon_n [1 - Z_{i \alpha}(\I \epsilon_n)] \mat 1 + \phi_{i \alpha}(\I \epsilon_n) \mat \sigma_1 + \chi_{i \alpha}(\I \epsilon_n) \mat \sigma_3,
\end{equation}
where $\mat \sigma_1$ and $\mat \sigma_3$ are Pauli matrices.
Correspondingly, the bare and screened electron Green functions become
\begin{multline}
    \label{eq:G2x2}
    \mat G_{i \alpha}(\I \epsilon_n) = -\int_{-\infty}^\infty \D \epsilon \, N_\alpha(\epsilon)
    \\
    \times \frac{\I \epsilon_n Z_{i \alpha}(\I \epsilon_n) \mat 1 + \phi_{i \alpha}(\I \epsilon_n) \mat \sigma_1 + [\epsilon - \mu_i + \chi_{i \alpha}(\I \epsilon_n)] \mat \sigma_3}{\Theta_{i \alpha}(\epsilon, \I \epsilon_n)},
\end{multline}
with the noninteracting values $Z_{0 \alpha}(\I \epsilon_n) = 1$, $\phi_{0 \alpha}(\I \epsilon_n) = 0$, and $\chi_{0 \alpha}(\I \epsilon_n) = 0$.
$\Theta_{i \alpha}(\epsilon, \I \epsilon_n)$ is the function whose zeros define the quasiparticle excitation energies and is given by
\begin{equation}
    \Theta_{i \alpha}(\epsilon, \I \epsilon_n) = [\epsilon_n Z_{i \alpha}(\I \epsilon_n)]^2
    + \phi_{i \alpha}^2(\I \epsilon_n) + [\epsilon - \mu_i + \chi_{i \alpha}(\I \epsilon_n)]^2.
\end{equation}
The self-energy components satisfy the local Eliashberg equations
\begin{subequations}
\label{eq:Eliashberg}
\begin{align}
    \label{eq:Z}
    Z_{i \alpha}(\I \epsilon_m) &= 1 - \frac{k \sub B T}{\epsilon_m} \sum_\beta \int_{-\infty}^\infty \D \epsilon \, N_\beta(\epsilon) \sum_n \frac{\epsilon_n Z_{i - 1 \beta}(\I \epsilon_n)}{\Theta_{i - 1 \beta}(\epsilon, \I \epsilon_n)} \notag
    \\
    &\quad \times \tilde W_{\alpha \beta}(\I \epsilon_n - \I \epsilon_m),
    \\
    \phi_{i \alpha}(\I \epsilon_m) &= -k \sub B T \sum_\beta \int_{-\infty}^\infty \D \epsilon \, N_\beta(\epsilon) \sum_n \frac{\phi_{i - 1 \beta}(\I \epsilon_n)}{\Theta_{i - 1 \beta}(\epsilon, \I \epsilon_n)} \notag
    \\
    &\quad \times \tilde W_{\alpha \beta}(\I \epsilon_n - \I \epsilon_m),
    \\
    \label{eq:chi}
    \chi_{i \alpha}(\I \epsilon_m) &= k \sub B T \sum_\beta \int_{-\infty}^\infty \D \epsilon \, N_\beta(\epsilon) \sum_n \frac{\epsilon - \mu_{i - 1} + \chi_{i - 1 \beta}(\I \epsilon_n)}{\Theta_{i - 1 \beta}(\epsilon, \I \epsilon_n)} \notag
    \\
    &\quad \times \tilde W_{\alpha \beta}(\I \epsilon_n - \I \epsilon_m).
\end{align}
\end{subequations}
Also here, we conserve the electron number
\begin{equation}
    n \sub e = \sum_\beta \int_{-\infty}^\infty \D \epsilon \, N_\beta(\epsilon) \Biggl[ 1 - 2 k \sub B T \sum_n \frac{\epsilon - \mu_i + \chi_{i \beta}(\I \epsilon_n)}{\Theta_{i \beta}(\epsilon, \I \epsilon_n)} \Biggr].
\end{equation}
If we enforce the normal state by setting $\phi_{i \alpha}(\I \epsilon_n) = 0$, Eqs.~\eqref{eq:Sigma} and \eqref{eq:Eliashberg} are equivalent and yield solutions that are related by analytic continuation (see Appendix~\ref{app:imag_to_real}).
Here, we use Pad\'e approximants \cite{Vidberg1977} to go from the imaginary to the real axis.

Finally, the renormalization function evaluated at the Fermi level can also be expressed as \cite{Eliashberg1962}
\begin{equation}
    \label{eq:Z0}
    Z_{i \alpha}(0) = 1 - \frac{\partial \Re \Sigma_{i \alpha}(\epsilon)}{\partial \epsilon} \biggr|_{\epsilon = 0}.
\end{equation}
This mass-enhancement factor is the \emph{reciprocal} of the quasiparticle weight often denoted with the same letter.

\subsection{Momentum dependence and transport calculations}

The Green and spectral functions in Eq.~\eqref{eq:G} are local quantities where the $\vec k$ dependence has been averaged out.
However, we are interested in renormalized bands with $\vec k$ dependence.
We thus consider the inverse Green function
\begin{equation}
    \label{eq:Ginvk}
    G^{-1}_{i \vec k \alpha \beta}(\epsilon) = [\epsilon + \mu_i - \Sigma_{i \alpha}(\epsilon)] \delta_{\alpha \beta} - H_{\vec k \alpha \beta},
\end{equation}
where $H_{\vec k \alpha \beta}$ is the Kohn--Sham Hamiltonian in the orbital basis \footnote{For simplicity, we use $\mu_i$ calculated for $A_{i \alpha}(\epsilon)$ from Eq.~\eqref{eq:A} using Eq.~\eqref{eq:ne}, and we do not recalculate it for $A_{i \vec k}(\epsilon)$ from Eq.~\eqref{eq:Ak}.}.
This matrix in $\alpha, \beta$ can be inverted and transformed into the basis of the noninteracting bands,
\begin{equation}
    G_{i \vec k m n}(\epsilon) = \sum_{\alpha \beta} \bracket{\vec k m}{\vec k \alpha} G_{i \vec k \alpha \beta}(\epsilon) \bracket{\vec k \beta}{\vec k n}.
\end{equation}
The diagonal Green function $G_{i \vec k n} \equiv G_{i \vec k n n}$ and the band-resolved and total spectral functions are related by
\begin{subequations}
\begin{align}
    \label{eq:Gk}
    G_{i \vec k n}(\epsilon) &= \int_{-\infty}^\infty \D \epsilon' \frac{A_{i \vec k n}(\epsilon')}{\epsilon - \epsilon'},
    \\
    A_{i \vec k n}(\epsilon) &= -\frac 1 \pi \Im G_{i \vec k n}(\epsilon + \I \eta),
    \\
    \label{eq:Ak}
    A_{i \vec k}(\epsilon) &= \sum_n A_{i \vec k n}(\epsilon). \vphantom \int
\end{align}
\end{subequations}

We use the bubble approximation to the Green--Kubo formula for the optical conductivity \cite{Haule2010, Pickem2022, Abramovitch2023, Abramovitch2024, Lihm2025, Lihm2026}.
The real part reads
\begin{multline}
    \label{eq:sigmaOmega}
    \Re \sigma_{i x y}(\omega) = \frac{\pi \hbar e^2}{V \super{uc}} \frac 2 {N_{\vec k}} \sum_{\vec k n} \int_{-\infty}^\infty \D \epsilon \frac {f(\epsilon) - f(\epsilon + \omega)} \omega
    \\
    \times v_{\vec k n x} v_{\vec k n \smash[b] y} A_{i \vec k n}(\epsilon) A_{i \vec k n}(\epsilon + \omega),
\end{multline}
where $\vec v_{\vec k n} = \D \epsilon_{\vec k n} / \D (\hbar \vec k)$ is the band velocity, $e$ is the elementary charge, $V \super{uc}$ is the unit-cell volume, and the factor of two accounts for the twofold degeneracy of the bands.
The imaginary part is obtained via the Kramers--Kronig relation.
See Appendix~\ref{app:current_current_correlation} for the derivation.
In the $\omega \rightarrow 0$ limit, this yields the \ac{DC} conductivity
\begin{equation}
    \label{eq:sigma}
    \sigma_{i x y}(0) = \frac{\pi \hbar e^2}{V \super{uc}} \frac 2 {N_{\vec k}} \sum_{\vec k n} \int_{-\infty}^\infty \D \epsilon \, \delta(\epsilon) v_{\vec k n x} v_{\vec k n \smash[b] y} A_{i \vec k n}^2(\epsilon),
\end{equation}
where $\delta(\epsilon) = -\D f / \D \epsilon$ is a temperature-dependent bell-shaped function that limits the integral to the small energies of thermal excitations.
The resistivity is calculated as the inverse matrix $\mat \rho_i = \mat \sigma_i^{-1}(0)$, where we are only interested in the in-plane component $\rho_{i 1 1} = \rho_{i 2 2} \neq \rho_{i 3 3}$ for this work.

\section{Computational details}
\label{sec:implementation}

\begin{figure*}
    \includegraphics{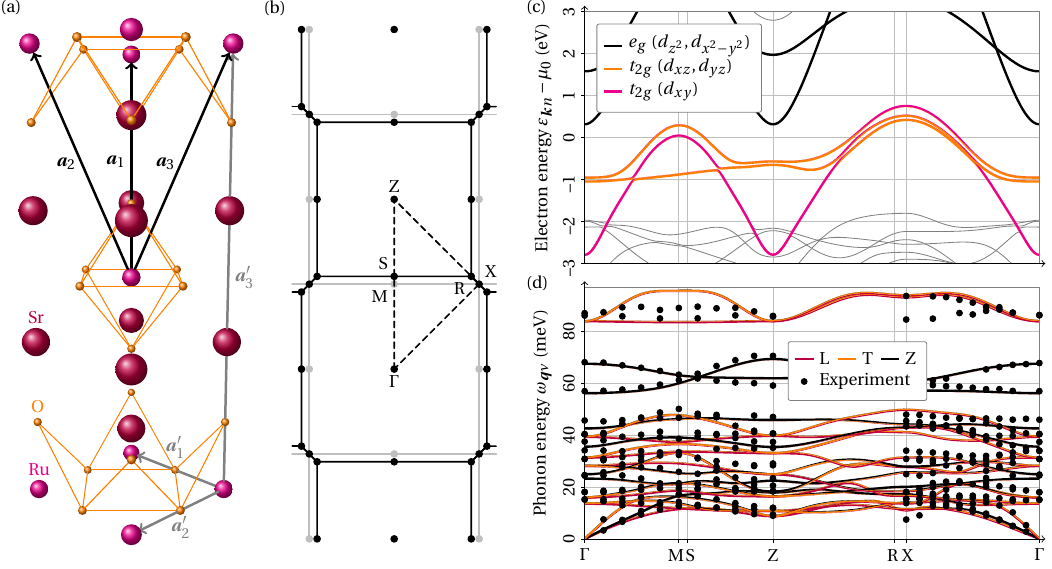}
    \caption{%
        (a)~Crystal structure and (b)~reciprocal space of Sr\s2RuO\s4.
        The primitive translation vectors, Brillouin-zone boundaries, and special $\vec k$~points of the simple (body-centered) tetragonal lattice are shown in gray (black).
        Note that the labels for the M and X points are often swapped in the case of the simple tetragonal lattice (see Fig.~4 of Ref.~\citenum{Setyawan2010}).
        (c)~Electron and (d)~phonon dispersion from \ac{DFT} and \ac{DFPT} along the path drawn with dashed lines in (b).
        The low-energy bands dominated by the $d$ orbitals of the Ru atoms are shown as thick lines; the $t_{2g}$ bands have been Wannier interpolated.
        Orbital characters and longitudinal (L), transverse (T), and out-of-plane (Z) phonon displacements are indicated with colors.
        Experimental phonon energies have been taken from \citeauthor{Braden2007} \cite{Braden2007}.
    }
    \label{fig:structure}
\end{figure*}

We calculate the electronic energies $\epsilon_{\vec k n}$ from \ac{DFT} and the phonon energies $\omega_{\vec q \nu}$ and electron--phonon coupling $g_{\vec q \nu \vec k \alpha \beta}$ from \ac{DFPT} using \textsc{Quantum ESPRESSO} \cite{Giannozzi2009, Giannozzi2017, Giannozzi2020} as well as the Coulomb interaction $W_{\alpha \beta}(\omega)$ from \ac{RPA} using \textsc{RESPACK} \cite{Nakamura2021}.
Here we use the \ac{PBE} functional \cite{Perdew1996} and norm-conserving pseudopotentials from the \textsc{PseudoDojo} table \cite{Hamann2013, vanSetten2018} at an energy cutoff of 90\,Ry.
We combine a cold smearing \cite{Marzari1999} of 20\,mRy with a $10 \times 10 \times 10$ $\vec k$-point sampling (including $\Gamma$) of the Brillouin zone of the body-centered tetragonal Bravais lattice of Sr\s2RuO\s4 [see black arrows and lines in Fig.~\ref{fig:structure}\,(a,\,b)].
Forces are minimized to below 1\,\textmu Ry/bohr at the experimental values of the lattice constants $a = 3.863$\,\AA{} and $c = 12.724$\,\AA{} \cite{Vogt1995}.
Phonons are computed on a $5 \times 5 \times 5$ $\vec q$-point mesh.
The \ac{RPA} calculation is performed considering 200 electronic bands and using a broadening of 0.01\,eV for the dynamical polarizability.
\Ac{SOC} is considered in the electronic energies but ignored in the interactions.

All quantities are transformed into the localized basis of one-shot Wannier functions based on atomic orbitals and atomic displacements using \textsc{Wannier90} \cite{Mostofi2008, Pizzi2020}, \textsc{EPW} \cite{Noffsinger2010, Ponce2016, Lee2023}, and \textsc{wan2respack} \cite{Kurita2023}.
On that basis, using the \textsc{elphmod} package \cite{Berges2017}, the local propagators as well as the spectral function and resistivity for a given self-energy are calculated using denser $\vec k$- and $\vec q$-point samplings generated through Fourier interpolation.
$N_\alpha(\epsilon)$ and $B \ph_{\alpha \beta}(\omega)$ are computed for a $\delta$-function smearing of 300\,K and 0.5\,meV in Eqs.~\eqref{eq:DOS} and \eqref{eq:Bph}, respectively.
In the phononic energy range, $B_{\alpha \beta}(\omega)$ is resampled on the logarithmic grid of $B \ph_{\alpha \beta}(\omega)$ using linear interpolation.
Here, $B_{\alpha \beta}(0) = 0$ is ensured by patching \textsc{RESPACK} to directly compute the retarded instead of time-ordered interaction.

The self-energy calculations are done using our open-source Eliashberg solver \textsc{ebmb} \cite{Berges2016a, Berges2016b}.
It is a Fortran program that calculates the band- or orbital-resolved local dynamical self-energy $G_i \tilde W_0$.
Here, we use $G_0 \tilde W_0$ but discuss $G \tilde W_0$ in Appendix~\ref{app:self-consistency}.
Calculations can be performed on the imaginary axis using Eq.~\eqref{eq:Eliashberg}, followed by analytical continuation to the real axis via Pad\'e approximants \cite{Vidberg1977}.
Normal-state results can also be obtained directly on the real axis, with logarithmic sampling to improve performance, as done in this work.
Here, we resort to the efficient Eqs.~\eqref{eq:ImSigma} and \eqref{eq:krakro}, where $A_{i \beta}(\epsilon \pm \omega)$ is derived from $A_{i \beta}(\epsilon)$ sampled on a fixed $\epsilon$ grid via linear interpolation.
We have verified that the two implementations yield consistent results (see Appendix~\ref{app:real_vs_imag}).
An OpenMP parallelization is used for the time-consuming computations of energy summations and integrals.
There is a Python wrapper that also provides some model inputs for testing, as used in Appendix~\ref{app:1D}.

\section{Study of Sr\s2RuO\s4}
\label{sec:results}

A ball-and-stick model of the Sr\s2RuO\s4 crystal is shown in Fig.~\ref{fig:structure}\,(a).
Its perovskite structure resembles that of the cuprate superconductor La\s2CuO\s4 \cite{Bauer2009}.
The conventional simple tetragonal unit cell, with the orthogonal primitive translation vectors shown in gray, contains two formula units.
However, we perform all calculations using the smaller primitive cell of the body-centered tetragonal lattice spanned by the vectors shown in black.
The corresponding reciprocal unit cells are shown in Fig.~\ref{fig:structure}\,(b) using the same shades.
Traversing the figure in a horizontal or vertical direction starting from $\Gamma$, in the case of the body-centered lattice, one passes from the middle plane of the first Brillouin zone to the floor or roof plane of the neighboring zone, while in the case of the simple tetragonal lattice, the vertical position within the cell stays the same.
The path $\Gamma$--M--S--Z--R--X--$\Gamma$ \cite{Boehnke2018} forms a right-angled triangle crossing the Brillouin-zone boundary.

The electronic band structure from \ac{DFT} along this path is shown in Fig.~\ref{fig:structure}\,(c).
Although we performed a noncollinear calculation to consider the effect of \ac{SOC}, because of the inversion symmetry of the crystal, all electronic bands remain twofold degenerate.
The relevant three low-energy bands have Ru-$t_{2g}$ ($d_{x z}$, $d_{y z}$, $d_{x y}$) orbital character.
The Ru-$e_g$ ($d_{z^2}$, $d_{x^2 - y^2}$) orbitals span the two conduction bands above.
In both cases, there is a significant admixture of O-$p$ orbital character.
In this work, we focus on the $t_{2g}$ bands, which we represent in the basis of one-shot Wannier functions based on projections onto the respective real $d$ orbitals, with an additional spin degree of freedom when considering \ac{SOC}.
In Fig.~\ref{fig:structure}\,(d), we show the Fourier-interpolated phonon dispersion from \ac{DFPT} together with results from inelastic neutron scattering \cite{Braden2007}.
Overall, the theoretical and experimental data match quite well, the notable exception being the transverse--optical mode at the M point and the Kohn anomaly at the X point.
Compared to previous calculations \cite{Braden1998, Bauer2009}, we do not observe any softening, most likely because of the high smearing employed.

\begin{figure*}
    \includegraphics{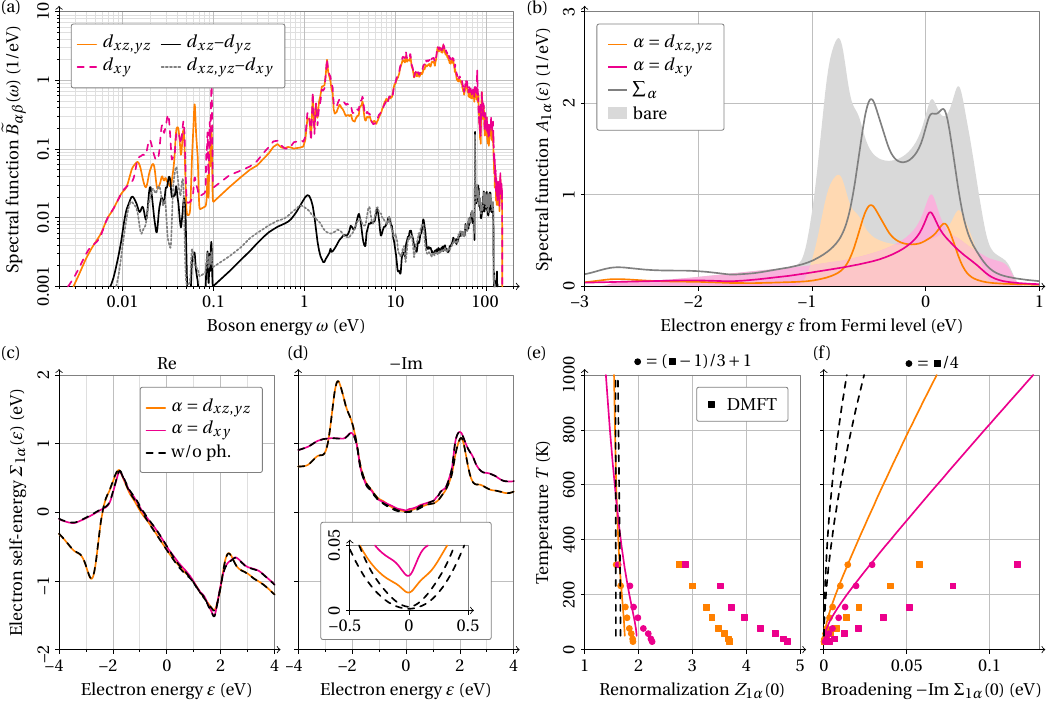}
    \caption{%
        Spectral functions of (a)~the effective interaction and (b)~the bare and screened Green functions of the Ru-$t_{2g}$ manifold of Sr\s2RuO\s4 at $T = 300$\,K.
        (c,\,d)~Energy dependence of retarded electron self-energy at $T = 300$\,K with and without the phonon-mediated part of the interaction.
        The inset is a close-up of the low-energy behavior of the imaginary part.
        (e,\,f)~Temperature dependence of the renormalization factor and broadening at the Fermi level.
        The \ac{DMFT} results, excluding phonon-mediated interaction, are from \citeauthor{Hunter2023} \cite{Hunter2023}.
    }
    \label{fig:local}
\end{figure*}

The spectral function of the effective interaction defined in Eq.~\eqref{eq:B} is shown in Fig.~\ref{fig:local}\,(a) for all combinations of orbitals in a double-logarithmic coordinate system.
It consists of the phonon \ac{DOS} weighted by the average electron--phonon coupling squared from \ac{DFPT} [Eq.~\eqref{eq:Bph}] and the scaled imaginary part of the screened Coulomb interaction from \ac{RPA} [Eq.~\eqref{eq:Bel}].
All peaks below and above 100\,meV are of phononic and electronic origin, respectively.
The intra-orbital interaction, as well as the exchange with the $d_{x y}$ orbital, is the same for the $d_{x z}$ and $d_{y z}$ orbitals, leaving four distinct matrix elements.
The off-diagonal exchange elements are overall much smaller than the diagonal density--density terms.
The distinct elements of the bare interaction are $v_{x z, x z} = 10.5$\,eV, $v_{x y, x y} = 11.5$\,eV, $v_{x z, y z} = 322$\,meV, and $v_{x z, x y} = 344$\,meV.

Next, the spectral functions of the bare and screened Green functions, that is the noninteracting electronic \ac{DOS} from \ac{DFT} defined in Eq.~\eqref{eq:DOS} and the spectral function or interacting \ac{DOS} defined in Eq.~\eqref{eq:A} are shown in Fig.~\ref{fig:local}\,(b).
The quantities are identical for the $d_{x z}$ and $d_{y z}$ orbitals in each case.
It is evident that the self-energy arising from the effective interaction [Eq.~\eqref{eq:Sigma}] leads to a renormalization of the bandwidth, which is reduced by about half (cf.\@ Ref.~\citenum{Zhang2016}).
The height of the spectral function is not correspondingly increased because a large part of the spectral weight is transferred to lower and higher energies.
This is the usual behavior captured by the \ac{QPA}, as discussed in Appendix~\ref{app:QPA}.

The self-energy itself is shown in Fig.~\ref{fig:local}\,(c,\,d), with and without the phonon-mediated part of the interaction.
The overall shapes of the real and imaginary parts are determined by the \ac{RPA} interaction, while the phonon-related features are hardly visible on the electronvolt scale.
However, a closer look at low energies reveals that phonons dominate the imaginary part at the Fermi level (see inset).
In Fig.~\ref{fig:local}\,(e,\,f), we show the renormalization factor defined in Eq.~\eqref{eq:Z0} and the broadening at the Fermi level as a function of temperature.
Here, especially at low temperatures, it is important to take the limit $\eta \rightarrow 0$ as done in Eq.~\eqref{eq:ImSigma}, since $\eta$ sets a lower bound for the magnitude of the imaginary part.
Interestingly, while the \ac{RPA} interaction largely defines the magnitude of the renormalization, the phonon-mediated interaction is responsible for its slight temperature variation and more than 80\,\% of the broadening.
These are direct consequences of the distinct energy scales of the two dynamical interactions, as shown in Appendix~\ref{app:1D} for a one-dimensional model, where we analyze the influence of the effective coupling strength $\lambda$ and the effective boson energy $\av \omega$ as used in Eliashberg theory \cite{Allen1982} on the renormalization and broadening at the Fermi level.
While the \ac{RPA} interaction acts with a strength of $\lambda_\alpha = \sum_\beta N_\beta(\mu_0) [v_{\alpha \beta} - W_{\alpha \beta}(0)] = 4.6$ (9.0) for $\alpha = d_{x z, y z}$ ($d_{x y}$) at an energy of $\av \omega = 13$\,eV, the phonons only contribute $\lambda \ph_\alpha = 0.21$ (0.42) but are much more effective at $\av \omega \ph = 30$\,meV.
The combined interaction, for which $\tilde \lambda_\alpha = 4.8$ (9.4) and $\av{\tilde \omega} = 10$\,eV, yields qualitative behavior in agreement with \ac{DMFT} results without phonons \cite{Hunter2023}, but quantitatively, the self-energy effects are much reduced here.
In the cases of the imaginary part and the slope of the real part, our results are off by factors of about four and three.

\begin{figure*}
    \includegraphics{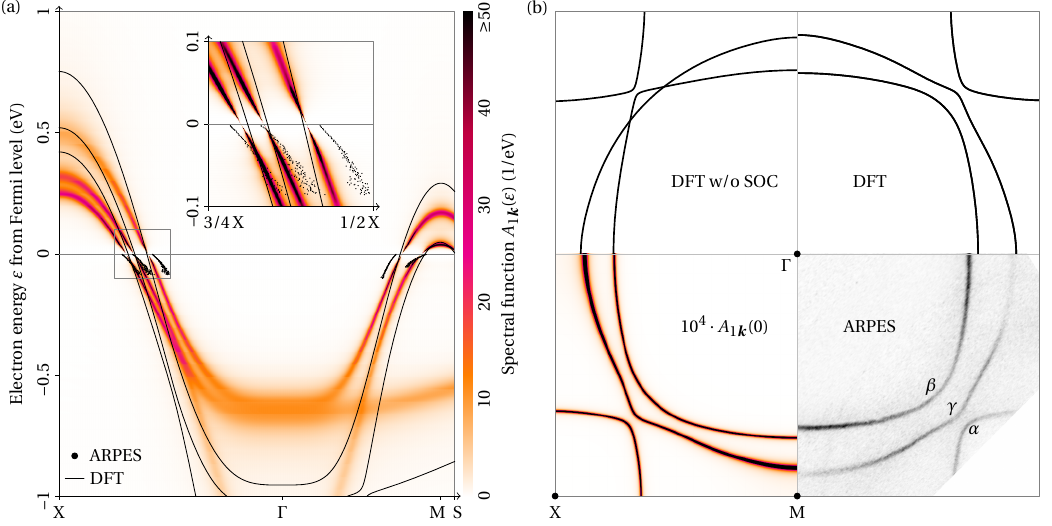}
    \caption{%
        Renormalized electronic structure of Sr\s2RuO\s4 at $T = 5$\,K compared to results from \ac{DFT} and \ac{ARPES} \cite{Tamai2019}.
        (a)~Spectral function.
        (b)~Fermi surface for $k_z = 0$.
        The $\beta$ and $\gamma$~sheets only cross in the absence of \ac{SOC}.
    }
    \label{fig:specfun}
\end{figure*}

We now complement the electronic bands of Fig.~\ref{fig:structure}\,(c) with the self-energy using Eq.~\eqref{eq:Ginvk}.
Note that \ac{SOC} requires an explicit spin dependence of the Kohn--Sham Hamiltonian $H_{\vec k \alpha \beta}$, so the number of Wannier orbitals $\alpha, \beta$ is doubled.
However, the same self-energy is added for both spin directions independently because the effect of \ac{SOC} only enters at the \ac{DFT} level; that is, it is already contained in $H_{\vec k \alpha \beta}$, and no additional \ac{SOC} term \cite{Tamai2019} is used.
Throughout the theory sections, we have assumed the electronic bands to be spin degenerate, so that the spin only manifests as simple factors of two.
Since a twofold degeneracy of all bands is preserved due to the inversion symmetry, all equations can be used as they are, even with \ac{SOC}.
The resulting $\vec k$-dependent low-temperature spectral function is shown in Fig.~\ref{fig:specfun}\,(a).
Reference values from \ac{ARPES} \cite{Tamai2019} are indicated with small dots.
The bands are significantly broadened, especially for larger energies; the broadening at the Fermi level depends on the temperature [see Fig.~\ref{fig:local}\,(e,\,f)].
The intensity maxima agree better with \ac{ARPES} than the bare \ac{DFT} bands, even though the match is not quantitative.
In the shown energy range, the renormalization strength is similar for all orbitals [see Fig.~\ref{fig:local}\,(c)] and its $\vec k$ and band dependence is correspondingly weak.
The inset is a close-up of the low-energy part of the spectrum, revealing slight phonon kinks with a reduction and enhancement of the band velocity at the Fermi level and further away, respectively.

The bottom--left quadrant of Fig.~\ref{fig:specfun}\,(b) shows the $\epsilon = 0$ cross section of the spectral function, that is, the Fermi surface, for $k_z = 0$.
The values are scaled by a factor of $10^4$ to be able to use the same color scale as in Fig.~\ref{fig:specfun}\,(a).
The other quadrants show corresponding results from \ac{DFT} with and without considering \ac{SOC} as well as \ac{ARPES} \cite{Tamai2019}.
The shape of the Fermi surface is most strongly influenced by \ac{SOC}, which lifts the crossing of the $\beta$ and $\gamma$ sheets, whereas the isotropic Hubbard corrections play a minor role \cite{Zhang2016}.
The main difference between the \ac{DFT} result with \ac{SOC} and our spectral function is the broadening.
Both match the measurement qualitatively, with some differences in the radial positions of the sheets.
Note that, especially on the $\Gamma$--M path, the $\beta$ sheet is less broadened than the $\gamma$ sheet and has a correspondingly increased intensity both in our calculation and in \ac{ARPES}.
This is because of the different orbital character ($d_{x z, y z}$ vs $d_{x y}$) of the sheets \cite{Iwasawa2010, Mravlje2011}, as seen in Figs.~\ref{fig:structure}\,(c) and \ref{fig:local}\,(f).

\begin{figure*}
    \includegraphics{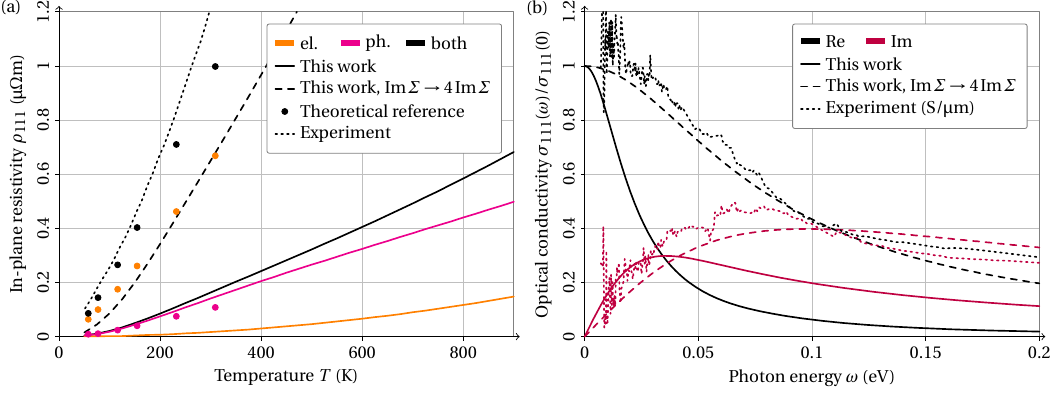}
    \caption{%
        Transport properties of Sr\s2RuO\s4.
        (a)~In-plane resistivity from the direct, phonon-mediated, and effective electron--electron scattering as a function of temperature.
        Theoretical reference data is from \citeauthor{Abramovitch2023} \cite{Abramovitch2023}, experimental data from \citeauthor{Tyler1998} \cite{Tyler1998}.
        (b)~Normalized in-plane optical conductivity from the effective interaction at $T = 290$\,K.
        Experimental data is from \citeauthor{Stricker2014} \cite{Stricker2014}.
    }
    \label{fig:resistivity}
\end{figure*}

Finally, we calculate the in-plane electrical resistivity according to Eq.~\eqref{eq:sigma} from the spectral function in Fig.~\ref{fig:specfun}.
The result is shown in Fig.~\ref{fig:resistivity}\,(a) as a function of temperature for the combined interaction as well as for the \ac{RPA} and phonon-mediated interactions separately.
The theoretical reference data from Ref.~\citenum{Abramovitch2023} is based on the same bubble-approximation formula given in Eq.~\eqref{eq:sigma}.
Note that off-diagonal velocities \cite{Pickem2022, Lihm2025, Lihm2026} have been omitted, which is justified because the bands are well separated near the Fermi level.
In the case of the phonon-mediated interaction alone, the only difference in the approaches is that we average the electron--phonon matrix elements from \ac{DFPT} to obtain a local interaction, while they calculate the Fan--Migdal self-energy with the full momentum and band dependence preserved.
Our data is close to the reference, especially at low temperatures.
The situation is different when considering the direct electron--electron interaction alone.
Here, their results are based on a spectral function from \ac{DMFT}, so the approaches differ fundamentally.
Using the dynamical Hubbard approach, we cannot reproduce their finding of a dominant role of the direct electron--electron interaction for the resistivity.
In fact, our values are not only smaller than theirs but even fall below the phonon-mediated contribution.
This is a consequence of the reduced broadening of our spectral function in the region of thermal excitations around the Fermi level.
Lastly, in the case of the combined interaction, they calculate the Fan--Migdal contribution using the Green function from \ac{DMFT} instead of the noninteracting bands from \ac{DFT}, while our one-shot local $G_0 \tilde W_0$ self-energy is based on the noninteracting \ac{DOS}.
Unlike theirs, our combined resistivity falls far below the experimental values measured with a four-probe transport technique \cite{Tyler1998}, which cannot solely be attributed to scattering channels ignored in our calculations.
Interestingly and as expected from the analysis in Fig.~\ref{fig:local}\,(f), scaling the imaginary part of the self-energy by a factor of four brings our results close to the \ac{DMFT} reference.
We only scale the imaginary part because the resistivity is approximately independent of the renormalization factor $Z_{1 \alpha}(0)$, as shown in Appendix~\ref{app:QPA}.
Similar observations can be made for the in-plane optical conductivity as per Eq.~\eqref{eq:sigmaOmega} and Appendix~\ref{app:current_current_correlation}, shown in Fig.~\ref{fig:resistivity}\,(b).
It decays faster than observed in optical reflectance measurements \cite{Stricker2014}, but scaling the linewidths by a factor of four improves the agreement.
We stress that vertex corrections to the transport equations, such as the self-consistent ladder approximation \cite{Lihm2026}, have been shown to produce very different transport results compared to the bubble approximation used here and in Ref.~\citenum{Abramovitch2023}, such that agreement with experiments is not necessarily expected.

\section{Conclusions}
\label{sec:conclusions}

We have demonstrated that a dynamical-Hubbard formulation that treats direct and phonon-mediated interactions consistently can, in the case of Sr\s2RuO\s4, yield self-energies and spectral functions that are qualitatively similar to those from \ac{DMFT}, at much reduced computational cost.
Especially the relative change of the self-energy with temperature is accurately described using our open-source implementation based on local Eliashberg theory \cite{Berges2016b}.
We have found that the bandwidth renormalization and the broadening at the Fermi level are largely determined by direct and phonon-mediated electron--electron interactions, respectively, which is a direct consequence of their very different effective coupling strengths $\lambda$ and energies $\av \omega$.
The broadening and thus the resistivity, which is almost independent of the renormalization factor, are underestimated significantly, which is likely related to the absence of vertex corrections \cite{Maebashi1998, Lihm2026} and Hund's coupling \cite{Mravlje2011} as well as overscreening in the \ac{RPA} \cite{Han2018, Honerkamp2018} and deserves further investigation.

\begin{acknowledgments}
    We acknowledge support from the \ac{DFG} under Germany's Excellence Strategy (University Allowance, EXC~2077, University of Bremen).
    J.B. acknowledges useful discussions with Jae-Mo Lihm and Manuel Zingl as well as the computing time granted by the Resource Allocation Board and provided on the supercomputers Lise and Emmy at NHR@ZIB and NHR@G\"ottingen as part of the NHR infrastructure.
    S.P. is a Research Associate of the \ac{FNRS}.
    This work was supported by the \ac{FNRS} under Grants number T.0183.23 (PDR) and T.W011.23 (PDR-WEAVE) and by the Walloon Region in the strategic axe FRFS-WEL-T.
\end{acknowledgments}

\section*{Data availability}

All code and data used are publicly available \cite{Berges2026}.

\appendix

\section{Temperature-dependent self-energy}
\label{app:matsubara}

Equilibrium many-body physics at finite electronic temperatures is properly described by the Matsubara formalism, where Green functions are evaluated on the imaginary axis and real-axis results follow from analytic continuation.
When evaluating diagrams, the dependence on electronic temperature or smearing enters via summations over Matsubara energies, while propagators do not explicitly depend on temperature.
This is different from the real-axis formalism, where the time-ordered propagators depend on temperature, while the integral expressions according to the Feynman rules do not \cite{Mattuck1967}.

Here we derive the expression for the self-energy in Eq.~\eqref{eq:Sigma}.
We consider fermionic and bosonic propagators of the form
\begin{align}
    \label{eq:fermionic_propagator}
    A(\I \epsilon_m) &= \int_{-\infty}^\infty \D \epsilon \frac{A(\epsilon)}{\I \epsilon_m - \epsilon},
    \\
    \label{eq:bosonic_propagator}
    B(\I \omega_n) &= B_0 + \int_{-\infty}^\infty \D \omega \frac{B(\omega)}{\I \omega_n - \omega},
\end{align}
where $B_0$ is a real constant.
Noting that $B(-\omega) = -B(\omega)$ allows us to rewrite the second equation with a positive range of integration as in Eqs.~\eqref{eq:Wph}, \eqref{eq:Wel}, and \eqref{eq:W}.
We are interested in the convolution
\begin{equation}
    \Sigma(\I \epsilon_m) = -k \sub B T \sum_n A(\I \epsilon_m + \I \omega_n) B(\I \omega_n),
\end{equation}
where $\omega_n = 2 n \pi k \sub B T$ with integer $n$ are bosonic Matsubara energies.
Inserting Eqs.~\eqref{eq:fermionic_propagator} and \eqref{eq:bosonic_propagator} gives
\begin{multline}
    \Sigma(\I \epsilon_m) = -k \sub B T \sum_n \int_{-\infty}^\infty \D \epsilon \frac{A(\epsilon)}{\I \epsilon_m + \I \omega_n - \epsilon}
    \\
    \times \Biggl[ B_0 + \int_{-\infty}^\infty \D \omega \frac{B(\omega)}{\I \omega_n - \omega} \Biggr].
\end{multline}
This can be recast as
\begin{multline}
    \Sigma(\I \epsilon_m) = -k \sub B T \Biggl[ B_0 \int_{-\infty}^\infty \D \epsilon \, A(\epsilon) \sum_n \frac 1 {\I \omega_n - (\epsilon - \I \epsilon_m)}
    \\
    + \iint_{-\infty}^\infty \D \epsilon \, \D \omega \, A(\epsilon) B(\omega) \sum_n \frac 1 {\I \omega_n - (\epsilon - \I \epsilon_m)} \frac 1 {\I \omega_n - \omega} \Biggr].
\end{multline}
Performing the Matsubara summations yields
\begin{multline}
    \Sigma(\I \epsilon_m) = B_0 \int_{-\infty}^\infty \D \epsilon \, A(\epsilon) \bigl( \tfrac12 + n(\epsilon - \I \epsilon_m) \bigr)
    \\
    + \iint_{-\infty}^\infty \D \epsilon \, \D \omega \, A(\epsilon) B(\omega) \frac{n(\omega) - n(\epsilon - \I \epsilon_m)}{\omega - (\epsilon - \I \epsilon_m)},
\end{multline}
where we have chosen to perform the first summation in an order where pairs $(-n, n)$ are summed first, so that it converges to a definite real value (cf.\@ Appendix~A of Ref.~\citenum{Allen1982}).
Using the relation $n(\epsilon \pm \I \epsilon_m) = -f(\epsilon)$ between the Bose and Fermi functions leads to
\begin{multline}
    \label{eq:GW_convolution}
    \Sigma(\I \epsilon_m) = B_0 \int_{-\infty}^\infty \D \epsilon \, A(\epsilon) \bigl( \tfrac12 - f(\epsilon) \bigr)
    \\
    + \iint_{-\infty}^\infty \D \epsilon \, \D \omega \, A(\epsilon) B(\omega) \frac{f(\epsilon) + n(\omega)}{\I \epsilon_m - \epsilon + \omega}.
\end{multline}
Analytic continuation $\I \epsilon_m \rightarrow \epsilon$, using $B(-\omega) = -B(\omega)$, and adding orbital indices yields Eq.~\eqref{eq:Sigma}.

Using $f(-\epsilon) = 1 - f(\epsilon)$ and $n(-\omega) = -1 - n(\omega)$, for $T \rightarrow 0$ with $\epsilon, \omega > 0$, we find $f(\epsilon) + n(\omega) \rightarrow 0^+$, $f(-\epsilon) + n(-\omega) \rightarrow 0^-$, $f(\epsilon) + n(-\omega) \rightarrow -1$, and $f(-\epsilon) + n(\omega) \rightarrow +1$.
Hence, significant contributions are only expected from poles $\epsilon$ and $\omega$ on different sides of the imaginary axis.
For $T = 0$, where the Matsubara summation becomes an integral over the imaginary axis, this immediately follows from Cauchy's residue theorem, as otherwise the contour could trivially be closed on the pole-less half-plane.
Similar arguments hold for poles on different sides of the real axis in the real-axis formulation.

Since the dynamical Hubbard approach usually relies on a real-axis formulation in terms of time-ordered \acp{SOP} \cite{Engel1993, Ferretti2024, Chiarotti2022, Chiarotti2023, Chiarotti2024, Chiarotti2025}, with poles located in opposite quadrants of the complex plane, we briefly compare the corresponding expressions to the above.
If we approximate the propagators in Eqs.~\eqref{eq:fermionic_propagator} and \eqref{eq:bosonic_propagator} by \acp{SOP} of the form
\begin{align}
    A(\epsilon) &= \sum_i \frac{a_i}{\epsilon - \alpha_i},
    \\
    B(\omega) &= B_0 + \sum_j \frac{b_j}{\omega - \beta_j},
\end{align}
with residues $a_i, b_j$ and \emph{complex} poles $\alpha_i, \beta_j$, the convolution yields, instead of Eq.~\eqref{eq:GW_convolution}, another \ac{SOP}
\begin{equation}
    \label{eq:convolution}
    \Sigma(\epsilon) = B_0 \sum_i a_i \bigl( \tfrac12 - f(\alpha_i) \bigr) + \sum_{i j} \frac{ a_i b_j [f(\alpha_i) + n(\beta_j)]}{\epsilon - (\alpha_i - \beta_j)}.
\end{equation}
This analytical expression for a \ac{SOP} resulting from a convolution of two \acp{SOP} is numerically efficient for a small number of poles.
However, a \ac{SOP} with few poles can only be a good approximation to a real-axis propagator if the poles lie away from the real axis, thus condensing many poles into few broadened ones.
After each convolution, the growing number of poles must be counteracted by an adequate condensation procedure.

At $T = 0$, Eq.~\eqref{eq:convolution} follows from Cauchy's residue theorem, and the occupation terms in brackets are reduced to one or zero, depending on whether the poles are located on different sides of the real axis.
In a temperature-dependent generalization, Fermi factors are usually incorporated into the electronic residues rather than the convolution itself \cite{Leon2021, Chiarotti2025}, neglecting the temperature dependence of the bosonic occupations.
While a \ac{SOP} can often well reproduce the overall shape of a propagator, difficulties may arise in the accurate description of low-temperature effects, as the number of poles in the low-energy region, where the Fermi factors are different from one or zero, is usually too small to capture relevant details.
Moreover, it is challenging to ensure that the imaginary parts of the time-ordered Green function and self-energy change sign exactly at the Fermi level throughout the calculation.
For these reasons, here we have chosen approaches based on Matsubara summation and numerical integration on a densely sampled real axis.

\section{Equivalence of formulations on different axes}
\label{app:imag_to_real}

\begin{figure*}
    \includegraphics{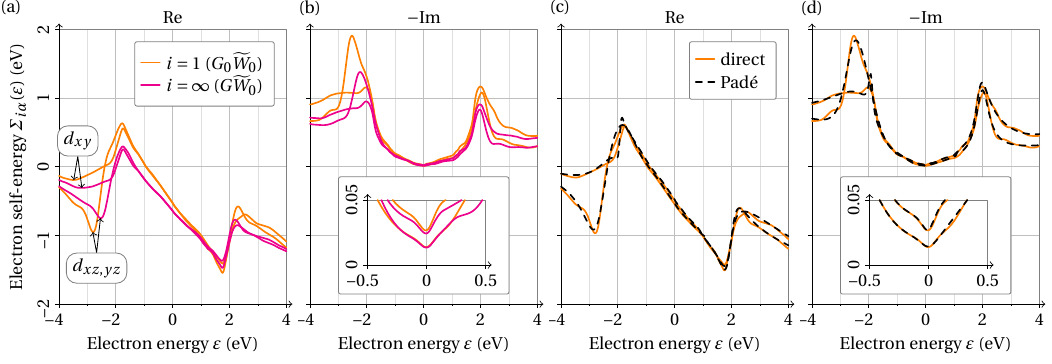}
    \caption{%
        Different ways of calculating the retarded electron self-energy of Sr\s2RuO\s4 at $T = 300$\,K.
        (a,\,b)~Comparison of one-shot local $G_0 \tilde W_0$ and self-consistent local $G \tilde W_0$ calculations on the real axis, enforcing $\Re \Sigma_{i \alpha}(0) = \Re \Sigma_{i \beta}(0)$ for all orbitals $\alpha, \beta$.
        (c,\,d)~Comparison of $G_0 \tilde W_0$ calculations directly on the real axis using Eqs.~\eqref{eq:ImSigma} and \eqref{eq:krakro} and on the imaginary axis using Eq.~\eqref{eq:Eliashberg} with subsequent analytic continuation via Pad\'e approximants.
    }
    \label{fig:variants}
\end{figure*}

The $(1, 1)$ component of Eq.~\eqref{eq:Sigma2x2} gives the normal-state self-energy
\begin{equation}
    \label{eq:Sigma11}
    \Sigma_{i \alpha}(\I \epsilon_n) = \I \epsilon_n [1 - Z_{i \alpha}(\I \epsilon_n)] + \chi_{i \alpha}(\I \epsilon_n).
\end{equation}
Equivalently, the $(1, 1)$ component of Eq.~\eqref{eq:G2x2} for $\phi_{i \alpha}(\I \epsilon_n) = 0$ gives the normal-state bare and screened Green functions
\begin{align}
    \label{eq:G11}
    G_{i \alpha}(\I \epsilon_n) &= -\int_{-\infty}^\infty \D \epsilon \, N_\alpha(\epsilon) \notag
    \\
    &\quad \times \frac{\I \epsilon_n Z_{i \alpha}(\I \epsilon_n) + \epsilon - \mu_i + \chi_{i \alpha}(\I \epsilon_n)}{[\epsilon_n Z_{i \alpha}(\I \epsilon_n)]^2 + [\epsilon - \mu_i + \chi_{i \alpha}(\I \epsilon_n)]^2}
    \\
    &= \int_{-\infty}^\infty \D \epsilon \frac{N_\alpha(\epsilon)}{\I \epsilon_n Z_{i \alpha}(\I \epsilon_n) - \epsilon + \mu_i - \chi_{i \alpha}(\I \epsilon_n)}.
\end{align}
Inserting Eqs.~\eqref{eq:Z} and \eqref{eq:chi} for $\phi_{i \alpha}(\I \epsilon_n) = 0$ into Eq.~\eqref{eq:Sigma11} and using Eq.~\eqref{eq:G11} yields
\begin{align}
    \Sigma_{i \alpha}(\I \epsilon_m) &= k \sub B T \sum_\beta \int_{-\infty}^\infty \D \epsilon \, N_\beta(\epsilon) \notag
    \\
    &\quad \times \sum_n \frac{\I \epsilon_n Z_{i - 1 \beta}(\I \epsilon_n) + \epsilon - \mu_{i - 1} + \chi_{i - 1 \beta}(\I \epsilon_n)}{[\epsilon_n Z_{i - 1 \beta}(\I \epsilon_n)]^2 + [\epsilon - \mu_{i - 1} + \chi_{i - 1 \beta}(\I \epsilon_n)]^2} \notag
    \\
    &\quad \times \tilde W_{\alpha \beta}(\I \epsilon_n - \I \epsilon_m)
    \\
    &= -k \sub B T \sum_{\beta n} G_{i - 1 \beta}(\I \epsilon_n) \tilde W_{\alpha \beta}(\I \epsilon_n - \I \epsilon_m)
    \\
    &= -k \sub B T \sum_{\beta n} G_{i - 1 \beta}(\I \epsilon_m + \I \omega_n) \tilde W_{\alpha \beta}(\I \omega_n).
\end{align}
Together with Appendix~\ref{app:matsubara}, this proves the equivalence of Eqs.~\eqref{eq:Sigma} and \eqref{eq:Eliashberg}.
Numerically, this is shown in Fig.~\ref{fig:variants}\,(c,\,d).

\section{Current--current correlation function}
\label{app:current_current_correlation}

The spatially integrated current--current correlation function in the independent-particle approximation can be written as \cite{Schweitzer1991, vandenBrink1998, Mahan2000, Benfatto2005, Tomczak2007, Ponce2020, Mu2022}
\begin{multline}
    \mathcal J_{x y}(\I \omega_l) = \frac{\hbar e^2}{V \super{uc}} \frac 2 {N_{\vec k}} \sum_{\vec k n} v_{\vec k n x} v_{\vec k n \smash[b] y}
    \\
    \times k \sub B T \sum_m G_{\vec k n}(\I \epsilon_m) G_{\vec k n}(\I \epsilon_m + \I \omega_l),
\end{multline}
where we have omitted the iteration index $i$ and the factor of two accounts for the twofold degeneracy of the bands.
With the spectral representation of the Green functions from Eq.~\eqref{eq:Gk}, this becomes
\begin{multline}
    \mathcal J_{x y}(\I \omega_l) = \frac{\hbar e^2}{V \super{uc}} \frac 2 {N_{\vec k}} \sum_{\vec k n} v_{\vec k n x} v_{\vec k n \smash[b] y}
    \\
    \times k \sub B T \sum_m \iint_{-\infty}^\infty \D \epsilon \, \D \epsilon' \frac{A_{\vec k n}(\epsilon)}{\I \epsilon_m - \epsilon} \frac{A_{\vec k n}(\epsilon')}{\I \epsilon_m + \I \omega_l - \epsilon'}.
\end{multline}
Performing the Matsubara summation yields
\begin{multline}
    \label{eq:J}
    \mathcal J_{x y}(\I \omega_l) = \frac{\hbar e^2}{V \super{uc}} \frac 2 {N_{\vec k}} \sum_{\vec k n} v_{\vec k n x} v_{\vec k n \smash[b] y}
    \\
    \times \iint_{-\infty}^\infty \D \epsilon \, \D \epsilon' \frac{f(\epsilon) - f(\epsilon')}{\I \omega_l + \epsilon - \epsilon'} A_{\vec k n}(\epsilon) A_{\vec k n}(\epsilon').
\end{multline}
Using $\Im[1 / (\epsilon + \I 0^+)] = -\pi \delta(\epsilon)$, the imaginary part of the retarded function follows as \cite{Haule2010}
\begin{multline}
    \label{eq:ImJ}
    \Im \mathcal J_{x y}(\omega + \I 0^+) = -\frac{\pi \hbar e^2}{V \super{uc}} \frac 2 {N_{\vec k}} \sum_{\vec k n} v_{\vec k n x} v_{\vec k n \smash[b] y}
    \\
    \times \int_{-\infty}^\infty \D \epsilon \, \bigl( f(\epsilon) - f(\epsilon + \omega) \bigr) A_{\vec k n}(\epsilon) A_{\vec k n}(\epsilon + \omega).
\end{multline}
The real part of the optical conductivity is defined as \cite{Tomczak2007}
\begin{equation}
    \Re \sigma_{x y}(\omega) = -\frac {\Im \mathcal J_{x y}(\omega + \I 0^+)} \omega,
\end{equation}
which, together with Eq.~\eqref{eq:ImJ}, yields Eq.~\eqref{eq:sigmaOmega}.
According to the Kramers--Kronig relations, the imaginary part can be obtained from the Hilbert transform,
\begin{align}
    \Im \sigma_{x y}(\omega) &= \frac 1 \pi \int_{-\infty}^\infty \D \omega' \frac{\Re \sigma_{x y}(\omega')}{\omega - \omega'}
    \\
    &= -\frac 1 \pi \int_{-\infty}^\infty \D \omega' \frac{\Im \mathcal J_{x y}(\omega' + \I 0^+)}{\omega' (\omega - \omega')}.
\end{align}
Inserting Eq.~\eqref{eq:ImJ} yields
\begin{multline}
    \Im \sigma_{x y}(\omega) = \frac{\hbar e^2}{V \super{uc}} \frac 2 {N_{\vec k}} \sum_{\vec k n} v_{\vec k n x} v_{\vec k n \smash[b] y}
    \\
    \times \iint_{-\infty}^\infty \D \epsilon \, \D \omega' \frac{ f(\epsilon) - f(\epsilon + \omega')}{\omega' (\omega - \omega')} A_{\vec k n}(\epsilon) A_{\vec k n}(\epsilon + \omega').
\end{multline}
With the substitution $\omega' \rightarrow \epsilon' - \epsilon$, this becomes
\begin{multline}
    \Im \sigma_{x y}(\omega) = \frac{\hbar e^2}{V \super{uc}} \frac 2 {N_{\vec k}} \sum_{\vec k n} v_{\vec k n x} v_{\vec k n \smash[b] y}
    \\
    \times \iint_{-\infty}^\infty \D \epsilon \, \D \epsilon' \frac{ f(\epsilon) - f(\epsilon')}{(\epsilon' - \epsilon) (\omega + \epsilon - \epsilon')} A_{\vec k n}(\epsilon) A_{\vec k n}(\epsilon').
\end{multline}
Noting that $\displaystyle \frac 1 {\epsilon' - \epsilon} = \frac 1 \omega \Biggl[ 1 - \frac{\omega + \epsilon - \epsilon'}{\epsilon - \epsilon'} \Biggr]$, we get
\begin{multline}
    \Im \sigma_{x y}(\omega) = \frac{\hbar e^2}{V \super{uc}} \frac 2 {N_{\vec k}} \sum_{\vec k n} v_{\vec k n x} v_{\vec k n \smash[b] y}
    \\
    \times \frac 1 \omega \iint_{-\infty}^\infty \D \epsilon \, \D \epsilon' \Biggl[ \frac{ f(\epsilon) - f(\epsilon')}{\omega + \epsilon - \epsilon'} - \frac{ f(\epsilon) - f(\epsilon')}{\epsilon - \epsilon'} \Biggr]
    \\
    \times A_{\vec k n}(\epsilon) A_{\vec k n}(\epsilon').
\end{multline}
Comparing this to Eq.~\eqref{eq:J}, we find the expression for the complex optical conductivity to be
\begin{equation}
    \sigma_{x y}(\omega) = \I \frac {\mathcal J_{x y}(\omega + \I 0^+) - \mathcal J_{x y}(0)} \omega,
\end{equation}
where the static limit $\mathcal J_{x y}(0)$ is real and covers the contribution of the diamagnetic current \cite{Benfatto2005}.

\section{Self-consistency}
\label{app:self-consistency}

In Fig.~\ref{fig:variants}\,(a,\,b), we compare the self-energies calculated with ($G \tilde W_0$) and without ($G_0 \tilde W_0$) self-consistency in the Green function.
Only here, we have enforced the otherwise only approximately fulfilled condition $\Re \Sigma_{i \alpha}(0) = \Re \Sigma_{i \beta}(0)$ for all orbitals $\alpha, \beta$, since otherwise the \ac{DOS} corresponding to the $d_{x z, y z}$ and $d_{x y}$ orbitals unphysically drift apart during the iterations toward self-consistency.
This behavior could also be counteracted by adjusting the $\omega \rightarrow \infty$ part of the \ac{RPA} interaction, which largely cancels its $\omega = 0$ value.
We find that self-consistency slightly reduces the bandwidth renormalization while hardly changing the imaginary part of the self-energy at the Fermi level.
Since the former does not significantly influence the resistivity (see Appendix~\ref{app:QPA}), we have chosen the $G_0 \tilde W_0$ approach for this work.

\section{Real- vs imaginary-axis implementation}
\label{app:real_vs_imag}

In Fig.~\ref{fig:variants}\,(c,\,d), we compare the self-energies calculated on the different axes of the complex plane.
All colored solid lines have been calculated on the real axis directly using Eqs.~\eqref{eq:ImSigma} and \eqref{eq:krakro}.
All black dashed lines have been calculated on the imaginary axis using Eq.~\eqref{eq:Eliashberg} and analytically continued to the real axis via Pad\'e approximants \cite{Vidberg1977}.
Both approaches yield almost identical results, except for some noise from analytic continuation.
The imaginary-axis approach is more efficient at high temperatures and convenient because we can choose the energy samplings for calculations and plotting independently.
However, the low-energy spectral functions used for the resistivity are best calculated on the real axis directly, because even the slightest noise can significantly distort these results and low temperatures require many Matsubara energies.
For consistency, we have only used the real-axis approach in this work, except for this comparison.

\section{One-dimensional model}
\label{app:1D}

To further investigate the role of the effective coupling strength and the effective boson energy, we consider a one-dimensional model, similar to the one we already studied in Ref.~\citenum{Berges2023}.
For the electron dispersion, we use the nearest-neighbor tight-binding model $\epsilon(k) = -2 t \cos(k)$ with the hopping parameter $t$, which leads to a noninteracting \ac{DOS}
\begin{equation}
    N(\epsilon) = \frac 1 {\pi \sqrt{(2 t)^2 - \epsilon^2}}
\end{equation}
defining our bare Green function via Eq.~\eqref{eq:G0}.
Similarly, for the boson dispersion, we use the nearest-neighbor mass--spring model $\omega(q) = 2 \av \omega \sin(q / 2)$ with the effective energy $\av \omega$ defined below.
Finally, we add an electron--boson coupling $g(q, k) = \I g_0 \sqrt{2 \av \omega / \omega(q)} [\sin(k) - \sin(k + q)]$, derived from Harrison's rule for the variation of the hopping with the bond length.
Taking the $k$ average of its squared modulus yields $g^2(q) = \omega(q) g_0^2 / \av \omega$.
This leads to the Eliashberg spectral function
\begin{equation}
    \alpha^2 F(\omega) = N(\mu_0) g^2(\omega) F(\omega) = \frac{\lambda \omega}{\pi \sqrt{(2 \av \omega)^2 - \omega^2}}
\end{equation}
defining an interaction $W(\omega)$ as in Eq.~\eqref{eq:Wel} by setting $v = 0$ and $B(\omega) = \alpha^2 F(\omega) / N(\mu_0)$.
Here, $F(\omega)$ is the boson \ac{DOS} of the same shape as $N(\epsilon)$ and
\begin{equation}
    \lambda = -N(\mu_0) W(0) = 2 \int_0^\infty \frac {\D \omega} \omega \alpha^2 F(\omega) = \frac{2 N(\mu_0) g_0^2}{\av \omega}
\end{equation}
is the dimensionless effective coupling strength.
The effective boson energy is defined as
\begin{equation}
    \av \omega = \exp \Biggl[ \frac 2 \lambda \int_0^\infty \frac {\D \omega} \omega \log(\omega) \alpha^2 F(\omega) \Biggr].
\end{equation}

\begin{figure*}
    \includegraphics{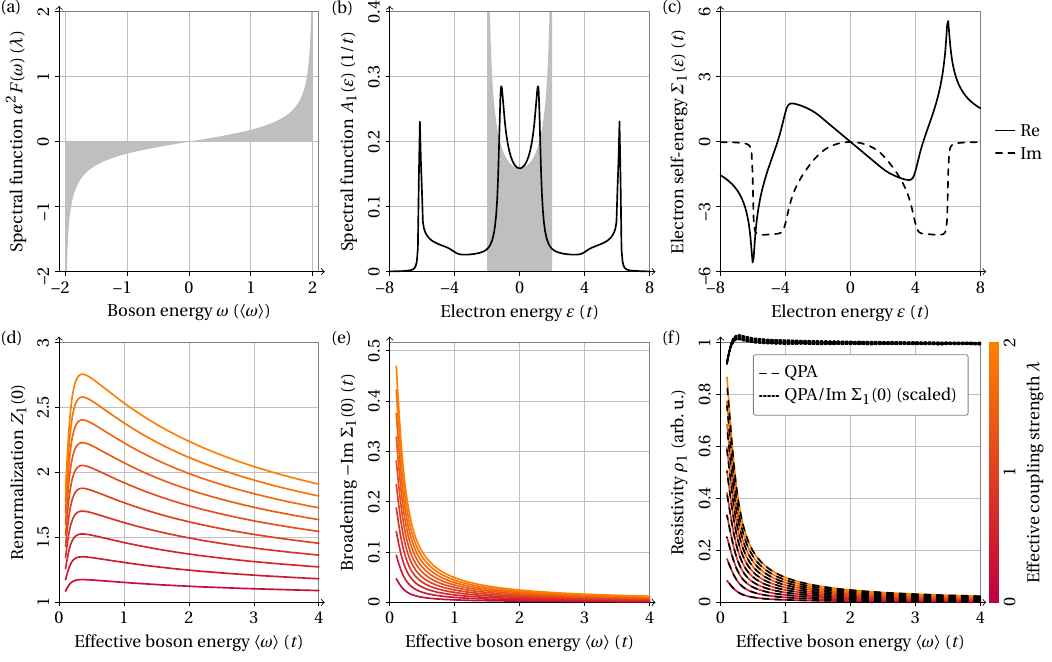}
    \caption{%
        Dynamical Hubbard approach for the one-dimensional model at $k \sub B T = 0.1 t$, where $t$ is the hopping parameter.
        (a)~Eliashberg spectral function.
        (b)~Bare and screened \ac{DOS}.
        (c)~Electron self-energy.
        (d)~Renormalization factor, (e)~broadening, and (f)~resistivity as a function of the effective boson energy for different effective coupling strengths.
        The \ac{QPA} results have been obtained using Eq.~\eqref{eq:conductivity_QPA}.
    }
    \label{fig:1D}
\end{figure*}

Figure~\ref{fig:1D} shows model results for half filling, that is $\mu_i = 0$.
The Eliashberg spectral function is shown in Fig.~\ref{fig:1D}\,(a).
It diverges for $\omega = 2 \av \omega$ and linearly crosses zero at $\omega = 0$.
The bare and screened electronic \ac{DOS} are shown in Fig.~\ref{fig:1D}\,(b).
The screening makes the peaks at $\pm 2 t$ approach each other, and satellites appear at higher energies (cf.\@ Ref.~\citenum{Aryasetiawan2022}).
The corresponding self-energy is shown in Fig.~\ref{fig:1D}\,(c).
The renormalization factor derived from its slope at zero energy is plotted in Fig.~\ref{fig:1D}\,(d) for different effective coupling strengths $\lambda$ and boson energies $\av \omega$.
While for small $\av \omega$ above the energy of thermal excitations it approaches $Z = 1 + \lambda$, it slowly decreases with increasing $\av \omega$.
The broadening at the Fermi level shown in Fig.~\ref{fig:1D}\,(e) also decreases with increasing $\av \omega$, albeit much faster.
This explains why the \ac{RPA} interaction of Sr\s2RuO\s4 can, at the same time, strongly renormalize the bands and only weakly broaden them near the Fermi level for small temperatures.
This equally affects the resistivity shown in Fig.~\ref{fig:1D}\,(f), whose $\av \omega$ dependence closely resembles that of the broadening in Fig.~\ref{fig:1D}\,(e), as explained in the following.

\section{Conductivity in the \acl*{QPA}}
\label{app:QPA}

Here, we analyze the dependence of the conductivity on the renormalization factor $Z_{\vec k n} = \sum_\alpha \abs{\bracket{\vec k n}{\vec k \alpha}}^2 Z_\alpha(0)$, where $Z_\alpha(0)$ is defined in Eq.~\eqref{eq:Z0}, in the \ac{QPA}.
We again omit the iteration index $i$.
It is important to note that for consistency with Eq.~\eqref{eq:Z}, our renormalization factor is the \emph{reciprocal} of what is often used in this context.
In Eliashberg theory and in this work, we consider the mass-enhancement factor $Z = 1 + \lambda = m^* / m$ (see Eq.~10 of Ref.~\citenum{McMillan1968}), while in $GW$ \cite{Leon2021} and \ac{DMFT} \cite{Mravlje2011, Deng2016}, we consider the quasiparticle weight $\mathcal Z = \epsilon / \epsilon_0 = m / m^*$ (see p.~545 of Ref.~\citenum{Mahan2000}).

Assuming $\mu = 0$ without loss of generality, the spectral function in the \ac{QPA} reads \cite{Martin2016}
\begin{equation}
    \label{eq:Adef}
    A \super{QPA}_{\vec k n}(\epsilon) \approx \frac 1 {\pi Z_{\vec k n}} \frac{\varGamma_{\vec k n}}{[\epsilon - E_{\vec k n}]^2 + \varGamma_{\vec k n}^2},
\end{equation}
where $E_{\vec k n} = \epsilon_{\vec k n} + \Re \Sigma(\epsilon_{\vec k n}) / Z_{\vec k n} \approx \epsilon_{\vec k n} / Z_{\vec k n}$ is the peak position and $\varGamma_{\vec k n} = -\Im \Sigma(E_{\vec k n}) / Z_{\vec k n}$ is the \ac{HWHM} of the Lorentzian line shape.
This spectral function integrates to the quasiparticle weight $1 / Z_{\vec k n} \leq 1$, and the missing weight $1 - 1 / Z_{\vec k n}$ belongs to the satellites \cite{Martin2016}.
Inserting Eq.~\eqref{eq:Adef} into Eq.~\eqref{eq:sigma} for the \ac{DC} conductivity yields, to leading order in the scattering rate $\varGamma_{\vec k n}$, the \ac{RTA} expression \cite{Pickem2022}
\begin{equation}
    \label{eq:conductivity_QPA}
    \sigma_{x y}(0) = \frac{e^2}{V \super{uc}} \frac 2 {N_{\vec k}} \sum_{\vec k n} \delta(E_{\vec k n}) V_{\vec k n x} V_{\vec k n \smash[b] y} \tau_{\vec k n},
\end{equation}
with the renormalized velocity $\vec V_{\vec k n} = \vec v_{\vec k n} / Z_{\vec k n}$ and the lifetime $\tau_{\vec k n} = \hbar / (2 \varGamma_{\vec k n})$.
Note that, unlike Eq.~\eqref{eq:sigma}, Eq.~\eqref{eq:conductivity_QPA} contains screened quantities only.
We can also make the $Z_{\vec k n}$ dependence explicit by writing
\begin{equation}
    \sigma_{x y}(0) = -\frac{\hbar e^2}{V \super{uc}} \frac 1 {N_{\vec k}} \sum_{\vec k n} \frac{\delta(\epsilon_{\vec k n} / Z_{\vec k n})}{Z_{\vec k n}} \frac{v_{\vec k n x} v_{\vec k n \smash[b] y}}{\Im \Sigma(\epsilon_{\vec k n} / Z_{\vec k n})}.
\end{equation}
The sum over the $\delta$ functions alone represents the density of renormalized energies at the Fermi level, broadened by temperature, which is proportional to $Z_{\vec k n}$.
The imaginary part of the self-energy is a quadratic function and is thus independent of $Z_{\vec k n}$ to first order.
Taken together, this shows that the resistivity is approximately independent of $Z_{\vec k n}$.

As shown in Fig.~\ref{fig:1D}\,(f), the resistivity from Eq.~\eqref{eq:sigma} is almost exactly reproduced by Eq.~\eqref{eq:conductivity_QPA}.
We find that the \ac{QPA} works similarly well for Sr\s2RuO\s4.
Indeed, if we divide the \ac{QPA} resistivity in Fig.~\ref{fig:1D}\,(f) by the imaginary part of the self-energy in Fig.~\ref{fig:1D}\,(e), we get a horizontal line for boson energies larger than the thermal excitations, proving the approximate independence of $Z_{\vec k n}$ usually expected from the resistivity \cite{Nakajima1963, Prange1964, Heine1966, Grimvall1976, Maebashi1997, Mahan2000, Chubukov2017} also numerically.

\bibliography{ms}

\end{document}